\documentclass[namedreferences,hyperref,optionalrh]{spr-sola}
\usepackage{graphicx}        
\usepackage{color}           

\usepackage{url} 
\usepackage{lineno}
\usepackage{soul}
\usepackage{amsmath} 
\usepackage{booktabs} 
\usepackage{natbib}
\usepackage{hyperref}
\hypersetup{linkcolor=blue,citecolor=cyan,filecolor=cyan,urlcolor=blue}
\usepackage{longtable}
\usepackage{caption}
\usepackage{enumitem}

\chardef\us=`\_

\begin{document}

\begin{frontmatter}
\title{SolARED: Solar Active Region Emergence Dataset for Machine Learning Aided Predictions}

\author[addressref={aff1,aff2},email={skasapis@princeton.edu}]{%
\inits{S.}\fnm{Spiridon}~\snm{Kasapis}}

\author[addressref=aff3,email={}]{%
\inits{E.}\fnm{Eren}~\snm{Dogan}}

\author[addressref=aff2,email={}]{%
\inits{I.N.}\fnm{Irina~N.}~\snm{Kitiashvili}}

\author[addressref={aff2,aff5},email={}]{%
\inits{A.G.}\fnm{Alexander~G.}~\snm{Kosovichev}}

\author[addressref=aff5,email={}]{%
\inits{J.T.}\fnm{John~T.}~\snm{Stefan}}

\author[addressref={aff2,aff4},email={}]{%
\inits{J.D.}\fnm{Jake~D.}~\snm{Butler}}

\author[addressref=aff3,email={}]{%
\inits{J.}\fnm{Jonas}~\snm{Tirona}}

\author[addressref=aff3,email={}]{%
\inits{S.}\fnm{Sarang}~\snm{Patil}}

\author[addressref=aff3,email={}]{%
\inits{M.}\fnm{Mengjia}~\snm{Xu}}

\address[id=aff1]{Department of Astrophysical Sciences, Princeton University, NJ, USA}

\address[id=aff2]{Computational Physics Branch, NASA Ames Research Center, Moffett Field, CA, USA}

\address[id=aff3]{Department of Data Science, New Jersey Institute of Technology, Newark, NJ, USA}

\address[id=aff5]{Center for Computational Heliophysics, Department of Physics, New Jersey Institute of Technology, Newark, NJ, USA}

\address[id=aff4]{InuTeq, LLC, MD, USA}

\runningauthor{Kasapis et al.}
\runningtitle{\textit{Solar Physics}}

\begin{abstract}
The development of accurate forecasts of solar eruptive activity has become increasingly important for preventing potential impacts on space technologies and exploration. Therefore, it is crucial to detect Active Regions (ARs) before they start forming on the solar surface. This will enable the development of early-warning capabilities for upcoming space weather disturbances. For this reason, we prepared the Solar Active Region Emergence Dataset (SolARED). The dataset is derived from full-disk maps of the Doppler velocity, magnetic field, and continuum intensity, obtained by the Helioseismic and Magnetic Imager (HMI) onboard the Solar Dynamics Observatory (SDO). SolARED includes time series of remapped, tracked, and binned data that characterize the evolution of acoustic power of solar oscillations, unsigned magnetic flux, and continuum intensity for 50 large ARs before, during, and after their emergence on the solar surface, as well as surrounding areas observed on the solar disc between 2010 and 2023. The resulting ML-ready SolARED dataset is designed to support enhancements of predictive capabilities, enabling the development of operational forecasts for the emergence of active regions. The SolARED dataset is available at \url{https://sun.njit.edu/sarportal/}, through an interactive visualization web application. 
\end{abstract}
\end{frontmatter}

%
%

\section{Introduction} \label{sec:Introduction}

The possibility of detection of large solar Active Regions (ARs) before they become visible using synoptic imaging of subsurface magnetic activity was demonstrated by \cite{ilonidis2011detection}, who detected strong acoustic travel-time anomalies of an order of 12 to 16 seconds as deep as 65~Mm beneath the solar surface. \cite{ilonidis2013helioseismic} proved that originating from deeper layers, these acoustic anomalies, associated with emerging active regions, rise to the surface at velocities up to 1~km/s. The deviations in the mean phase travel time of acoustic waves before the emergence of 46 large active regions were investigated by \cite{stefan2023exploring}, which showed the relationship between subsurface acoustic signals before the emergence and surface magnetic flux after the emergence. Another aspect of magnetic flux emergence was studied by \cite{attie2018precursors}, who observed disruption of the moat flow near active region AR12673, several hours before the onset of additional strong flux emergence events. This disruption occurred where magnetic flux later emerged, suggesting that horizontal divergent flows at the solar surface are potential precursors to this flux. In addition, divergent and convergent surface flows, which may serve as precursors to the emergence of ARs, have been extensively studied \citep[e.g.,][]{rees2022preemergence,gottschling2021evolution,schunker2024flux}. More recent studies focused on strengthening the solar surface gravity waves ($f$-mode) prior to emergence \citep{waidele2023strengthening,singh2016high}. 

While significant research has been devoted to understanding the complex behavior of subsurface activity preceding AR emergence, and despite the availability of relevant datasets \citep{schunker2016sdo}, only a few studies have directly addressed the challenge of predicting the emergence itself \citep{kasapis2023predicting,kasapis2025prediction,keegan2025data}. This gap in research primarily stems from the lack of publicly available datasets describing photospheric conditions before and during the emergence of ARs. 

The majority of space weather prediction models, especially those that utilize machine learning algorithms, have relied on publicly available datasets. For instance, \cite{bobra2021smarps} introduced the Space-Weather MDI Active Region Patches (SMARP) and Space-Weather HMI Active Region Patches (SHARP) data products that provide a continuous and seamless record of magnetic regions observed over the past two solar cycles, from 1996 to the present \citep{bobra2014helioseismic}, and have been used for a variety of space weather prediction applications such as solar flare forecasts \cite[e.g.,][]{sun2022predicting}. Similarly, studies that have used ML to predict SEP events \citep{kasapis2024forecasting, chatterjee2024mempsep} have trained their models on a variety of publicly available datasets \citep{moreland2024mempsep, hosseinzadeh2024improving, kosovich2024time} that are geared towards this specific application. Even datasets that correlate solar observations with in-situ measurements \citep{kasapis2023turning} have been developed to improve space weather forecasts, yet no dataset has been created specifically to predict the emergence of active regions. 

In this paper, we aim to fill this gap by providing the community with a data product derived from the Solar Dynamics Observatory’s Helioseismic and Magnetic Imager (SDO/HMI) \citep{scherrer2012helioseismic} called Solar Active Region Emergence Dataset (SolARED). SolARED includes time series that describe different physical quantities of 50 large ARs that emerged on the observed solar disc. Our approach involves local tracking and remapping onto Carrington heliographic coordinates of $30.66\times30.66$ heliographic-degree areas of magnetic field, continuum intensity, and Doppler velocity in the vicinity of selected ARs before, during, and after their emergence on the solar surface. SolARED includes time series of three different physical quantities: a) the continuum intensity, $I_c$; b) the unsigned magnetic flux, $\Phi_m,$; and c) the acoustic power, $P_a$. The $I_c$, $\Phi_m$, and $P_a$ (obtained from the Dopplergrams $V_d$) time series are derived by splitting the $30.66\times30.66$ degrees tracked patches in a 9-by-9 grid and then averaging the values within each grid tile. The dataset presented in this work aims to help understand the dynamics of emerging ARs and to aid with physics-based or machine-learning applications that predict the emergence and evolution of ARs. The steps followed to produce the dataset are outlined in Section \ref{sec:Processing}, including the tracking and remapping (to remove the solar sphere geometric effects) of the 50 ARs, the production of the acoustic power maps, and the reduction of the 3D data into time series apt for ML. The dataset contents are described in Section \ref{sec:DataPortal}, which presents the SolARED parameters for two illustrative ARs. Section~\ref{sec:Conclusions} concludes the paper with a discussion of the use cases of SolARED.

%
%

\section{Data Processing} \label{sec:Processing}

\begin{figure}[h] 
    \centering 
    \includegraphics[width=\textwidth]{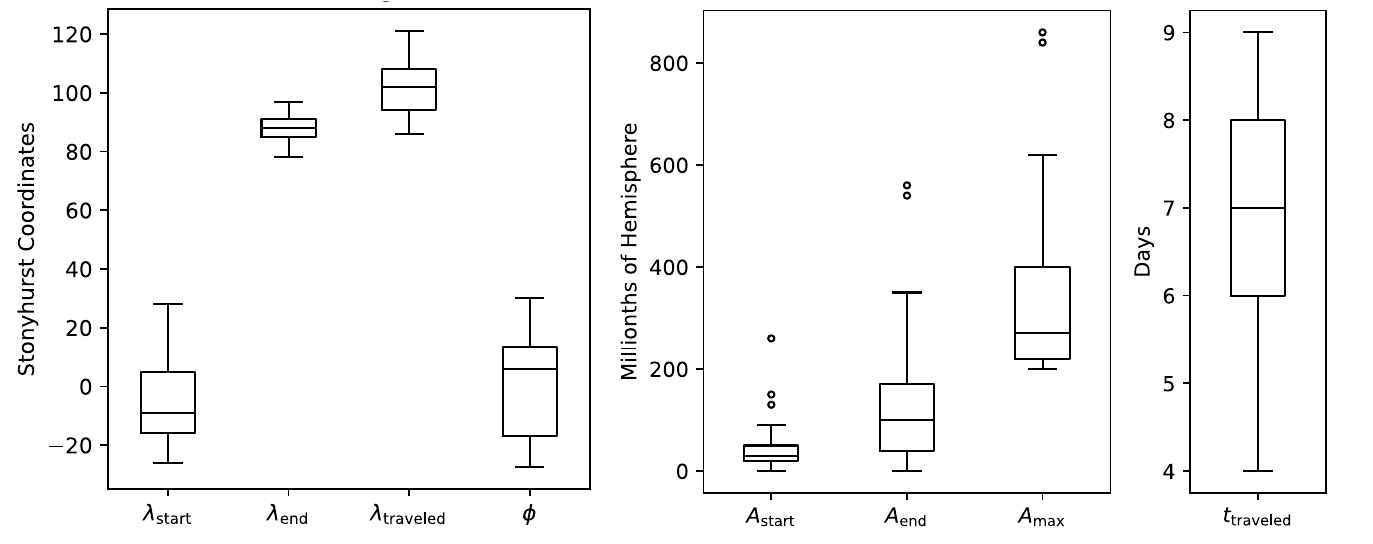} 
    \caption{Boxplots of the location, size, and observation time distributions for the 50 ARs included in SolARED. The left panel shows the distribution of starting ($\lambda_{start}$), ending ($\lambda_{end}$), and total traveled AR longitudes ($\lambda_{traveled}$) along with the constant latitude ($\phi$) in Stonyhurst coordinates. The middle panel shows the distribution of starting, ending, and maximum areas ($A_{start}, A_{end}$, and $A_{max}$) occupied by the ARs in millionths of a hemisphere. The right plot shows the distribution of the total time the ARs were visible on the disc($t_{traveled}$).} 
    \label{fig:ar_boxplots} 
\end{figure}

The process of creating SolARED began with identifying and creating a catalogue \citep[Appendix, Table 5 in][]{kasapis2025prediction} of ARs that emerged on the solar surface within 30 degrees longitude from the central meridian between March 1st, 2010 and June 1st, 2023, persisted for more than 4 days, and reached a total area of 200 millionths of the solar hemisphere. This longitude range was chosen to minimize significant distortion due to projection and center-to-limb effects. Figure~\ref{fig:ar_boxplots} presents the latitudinal and longitudinal distributions of emerging ARs along with statistics about their area growth and travel time from the East to the West limbs. More specifically, the median heliographic latitude of these 50 ARs is approximately $4^{\circ}$, with a standard deviation of about $17^{\circ}$, indicating a roughly symmetric distribution with respect to the solar equator. The selected ARs typically emerge around $-8^{\circ}$ Stonyhurst longitude, with approximately a quarter of them emerging in positive and the rest three-quarters emerging in negative longitudes. In terms of area evolution, the median starting, ending, and maximum areas are $30$, $110$, and $290$ millionths of a solar hemisphere (MH), respectively, with substantial variability (standard deviations of approximately $46$, $129$, and $173$) and two ARs growing significantly larger at $A_{max} > 800$ MH (AR 12085 and AR 13098). The median travel time of these 50 ARs on the visible solar surface is 7 days.

\begin{figure}[h] 
    \centering 
    \includegraphics[width=\textwidth]{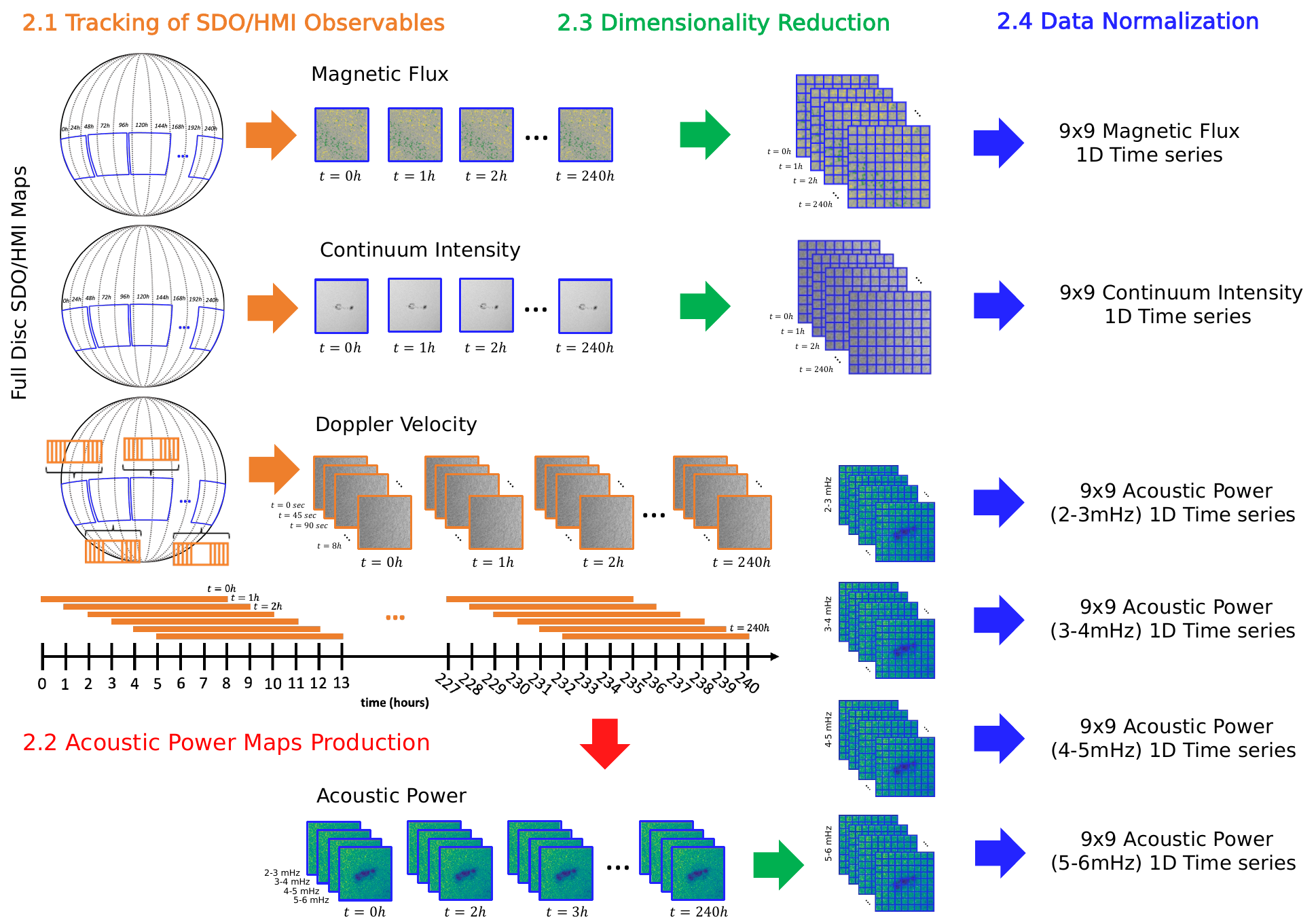} 
    \caption{The SDO/HMI observables processing pipeline used to create the SolARED dataset. The four different arrow colors (orange, red, green, and blue) represent the following processing steps: 1) tracking of SDO/HMI Observables, 2) production of acoustic power maps, 3) dimensionality reduction, and 4) data normalization, respectively. The blue rectangular regions on the spheres of Step 2.1 (Section~\ref{sec:tracking}) represent the $30.66^{\circ} \times 30.66^{\circ}$ ($512 \times 512$ pixels) areas that were tracked, while the blue grid in Step 2.3 (Section~\ref{sec:tiling}) is the 9 by 9 grid this area was split in.} 
    \label{fig:processing} 
\end{figure}

For each active region we performed the following processing steps (Figure~\ref{fig:processing}): a) tracking $30.66\times30.66$ degrees area with the local rotation rate for magnetic field, Doppler velocity, and continuum intensity (Section \ref{sec:tracking}); b) generating acoustic power maps from the Doppler velocity (Section \ref{sec:power_maps}), c) downsampling the data to timelines by splitting the tracked region in a 9 by 9 grid (Section \ref{sec:tiling}); and d) removing the solar sphere geometric effect (Section \ref{sec:flattening}).

\subsection{Tracking of SDO/HMI Observables} \label{sec:tracking}

The SDO/HMI observations of the 50 ARs in the SolARED catalog were tracked through time using $30.66^{\circ} \times 30.66^{\circ}$ patches representing a $512 \times 512$ pixel square centered around the middle of each AR. These patches were tracked before, during, and after the AR emergence, for a total of 10 days (240 hours). At each moment of time, the tracked area was remapped onto the heliographic coordinates using the equidistant azimuthal (Postel's) projection, which is typically used for local helioseismology analyses, the time-distance helioseismology \citep{zhao2012time-distance,bogart2011ring-diagram}. This remapping corrects the data for the geometrical effects. However, additional data normalization is required to correct for the line-of-sight projection of the Doppler velocity and magnetic field and for the limb darkening of the continuum intensity. This normalization is described in Section~\ref{fig:Correction}. 

As seen in Figure~\ref{fig:processing} (orange), a series of 240 SDO/HMI magnetograms ($M$) and continuum intensity ($I_c$) 2D maps, shifted by one hour during the AR tracking from the distance of about 75 degrees east to 75 degrees west from central meridian, is used to create the $M$ and $I_c$ timeseries. On the other hand, for the Dopplergrams $V_{\text{d}}$, 240 8-hour-long, one-hour-overlapping series of 2D maps were created, each containing 640 2D maps at a 45-second cadence. These 240 series of $V_{\text{d}}$ 2D maps (each series is represented by a $[640, 512, 512]$ data cube) are 1 hour apart from each other and are used to produce one acoustic power map per series (Figure~\ref{fig:processing}).


\subsection{Acoustic Power Maps Production} \label{sec:power_maps}

To generate power maps, we used the 240 8-hour Dopplergram time series $V_{\text{d}}$. To remove long-term variations on a scale of the tracked regions, for each series (640 frames with the 45-second cadence), we subtracted consecutive frames:
\begin{equation}
\Delta V_{\text{d}}(i,x,y) = V_{\text{d}}(i+1,x,y) - V_{\text{d}}(i,x,y), \quad \text{for } i = 1, \ldots, 639, \quad (x,y) \in [1,512].
\end{equation}
where $i$ is the temporal index that spans 8 hours, and $x$, $y$ are the spatial indices that correspond to the pixels within the tracked area. 

To compute the power spectral density maps, we applied 1D FFT to each pixel $(x, y)$ of the time-series data $\Delta V_{\text{d}}[:,x,y]$ for the 8-hour time series using the FFT Python library (\textit{np.fft.rfft}\footnote{\url{https://numpy.org/doc/2.3/reference/generated/numpy.fft.rfft.html}}):
\begin{equation}
\begin{split}
V_{\text{d}}^{\text{FFT}}(k,x,y) &= \left(\frac{dt^2}{T}\right) \left| \mathcal{F} \{ \Delta V_{\text{d}}(:,x,y) \} (k) \right|^2, \\
&\quad \text{for } k = 1, \ldots, 320,\; (x,y) \in [1,512].
\end{split}
\end{equation}
where $\mathcal{F}$ is the real part of the Fourier transform, $dt = 45$ sec (the sampling interval) and $T = 28800$ sec. 

The resulting power spectral density, $V_{\text{d}}^{\text{FFT}}$, array of size $[320, 512, 512]$ represents the Fourier power spectrum for each pixel $(x, y)$ over the selected frequency indices k. Given the total duration of the time series $T = 8 \times 60 \times 60 = 28{,}800$ seconds and the corresponding FFT indices $k = 0, 1, \ldots, N/2$, where $N = 640$ is the number of temporal frames, the physical frequency associated with each bin is computed as:
\begin{equation}
\nu(k) = \frac{k}{T} \times 1000,
\end{equation}
where $\nu(k)$ is the cyclic frequency expressed in mHz, though multiplying by a factor of 1000. 

For every $V_{\text{d}}^{\text{FFT}}(l)$ corresponding to a different .fits file, the integral is calculated along the frequency axis to construct a power map $P_{\text{map}}(l)$ of size $[512, 512]$. The power map is created based on a selected frequency range of interest. The analysis is concentrated on four frequency ranges: $2-3$, $3-4$, $4-5$, and $5-6$ mHz. The resulting $V_{\text{d}}^{\text{FFT}}$  was integrated over the selected frequency range for each pixel $(x, y)$ using the trapezoidal rule:
\begin{equation}
P(x,y) = \int_{\nu_{\text{lower}}}^{\nu_{\text{upper}}} V_{\text{d}}^{\text{FFT}}(\nu,x,y) \, d\nu, \quad \text{for } (x,y) \in [0,512],
\end{equation}
where $P[x,y]$ is the power map for each pixel $(x, y)$, which is a spatial map representing the integrated power spectral density within the selected frequency range [$\nu_{\text{lower}}, \nu_{\text{upper}}$]. By integrating over the frequencies in the selected range for each pixel, the power of the time series within the frequency range of interest is accumulated. This integration is performed numerically, resulting in a power map that represents the spectral power in the chosen frequency ranges at each spatial location on the tracked and remapped images.


\subsection{Dimensionality Reduction} \label{sec:tiling}

To efficiently train ML models, we reduced the data dimensionality to a 1D time series. The data reduction is performed by dividing the original tracked area of $30.66^{\circ} \times 30.66^{\circ}$ into a grid of 9-by-9 tiles, where each tile is $3.4^{\circ}\times3.4^{\circ}$ ($59 \times 59$ px per tile). 
By computing the mean acoustic power, continuum intensity, and unsigned magnetic flux for each tile in the maps, we compress the original dataset and filter out variations on small spatial scales that are not essential for predicting active region emergence. Each tile corresponds to an area of approximately $1.7\times10^3$ Mm$^2$ on the solar surface. 

\begin{figure}[h] 
    \centering 
    \includegraphics[width=\textwidth]{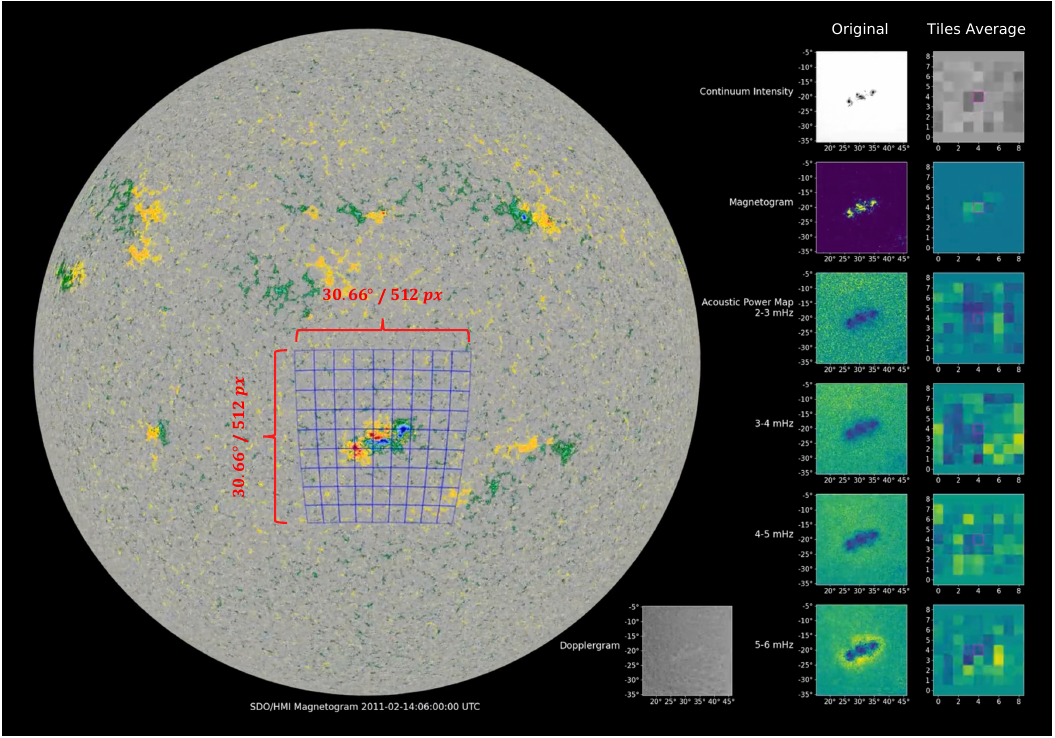} 
    \caption{Snapshot of the data tracking from a movie for the magnetic evolution of AR 11158 using the SDO/HMI observations. The panels at right show (from top to bottom) the evolution of the continuum intensity, the line-of-sight magnetic field, and the acoustic power in three frequency ranges. The two columns correspond to the original data of the tracked region (left) and the tile averages calculated after splitting the tracked area into a 9-by-9 grid (right).} 
    \label{fig:Video}
\end{figure}

Figure~\ref{fig:Video} illustrates the process\footnote{\url{https://www.nas.nasa.gov/SC24/research/project12.php\#demo}} of dimensionality reduction applied to the three tracked data products. The left panel shows a full-disk magnetogram, with the tracked region containing AR 11158 divided into a 9×9 grid. The right panels display snippets of this region (for all physical quantities used in this study), with the left column showing the original data and the right column showing the corresponding averaged values for each tile (Figure~\ref{fig:Correction}). 


\subsection{Data Normalization} \label{sec:flattening}

\begin{figure}[h] 
    \centering 
    \includegraphics[width=\textwidth]{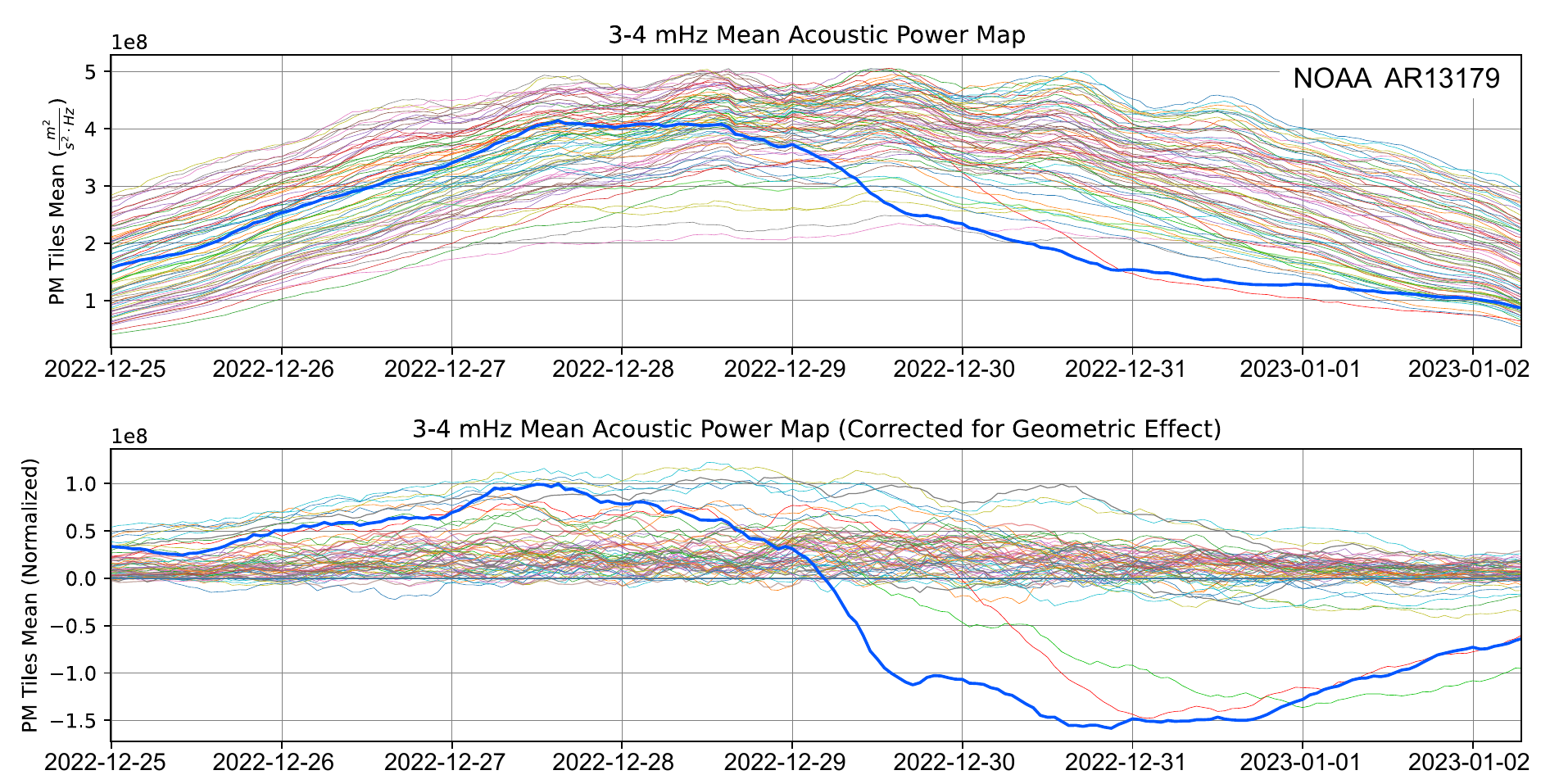} 
    \caption{The mean acoustic power time series for the 63 tiles of AR 13179, before (top) and after (bottom) the correction for the center-to-limb variations. The blue bold line represents a tile, where emergence is first observed.} 
    \label{fig:Correction} 
\end{figure}

The final step in the data processing pipeline is the removal of the center-to-limb effect that manifests as a systematic decrease in the continuum intensity and a change in the properties of the magnetic flux and power maps due to the line-of-sight projection and the increased formation height of the Fe\,I\,6173\,\AA\,\, line, utilized by the SDO/HMI instrument to obtain the corresponding observables. To remove these systematic variations (Figure~\ref{fig:Correction}, top panel), the data are normalized relative to the top and bottom row tiles for the same longitude (Figure \ref{fig:Video}, right column). Because these tiles typically correspond to quiet-Sun regions in the tracked ARs dataset, they are used as a baseline to normalize the intermediate-latitude tiles relative to zero. Normalization is performed along the longitudinal direction using the expression:
\begin{equation}
P(i) = P(i) - \left( \frac{i}{N} \cdot P(1) + \left( 1 - \frac{i}{N} \right) \cdot P(N) \right),
\end{equation}
where $P(i)$ are the values (of any physical quantity) that correspond to the $i^{th}$ row, $P(1)$ and $P(N)$ are the top and bottom rows, and $i = 1, 2, \dots, N$ ($N = 9$ for this study). 

This subtraction of temporal variations caused by the center-to-limb variations, combined with normalization to the quiet-Sun values, enhances the variations associated with magnetic activity (Figure~\ref{fig:Correction}, bottom panel). However, the reference tiles may contain some magnetic activity that could contribute to residual deviations. Given that the top and bottom rows of the 9-by-9 grid are normalized to zero, the total number of time series for each physical quantity and for each AR is 63 ($81 - 2 \times 9$). 


%
%

\section{SolARED Data Portal} \label{sec:DataPortal}

The ML-ready SolARED dataset is accessible through a web-based interactive visualization tool at \url{https://sun.njit.edu/sarportal/}, hosting the Intelligent Database of Solar Events and Active Regions (IDSEAR) portal at the New Jersey Institute of Technology. It offers a convenient way to plot, visualize, and download data to explore the time evolution of areas where magnetic activity develops, as well as regions where conditions remain quiet. 

\begin{figure}[h] 
    \centering 
    \includegraphics[width=\textwidth]{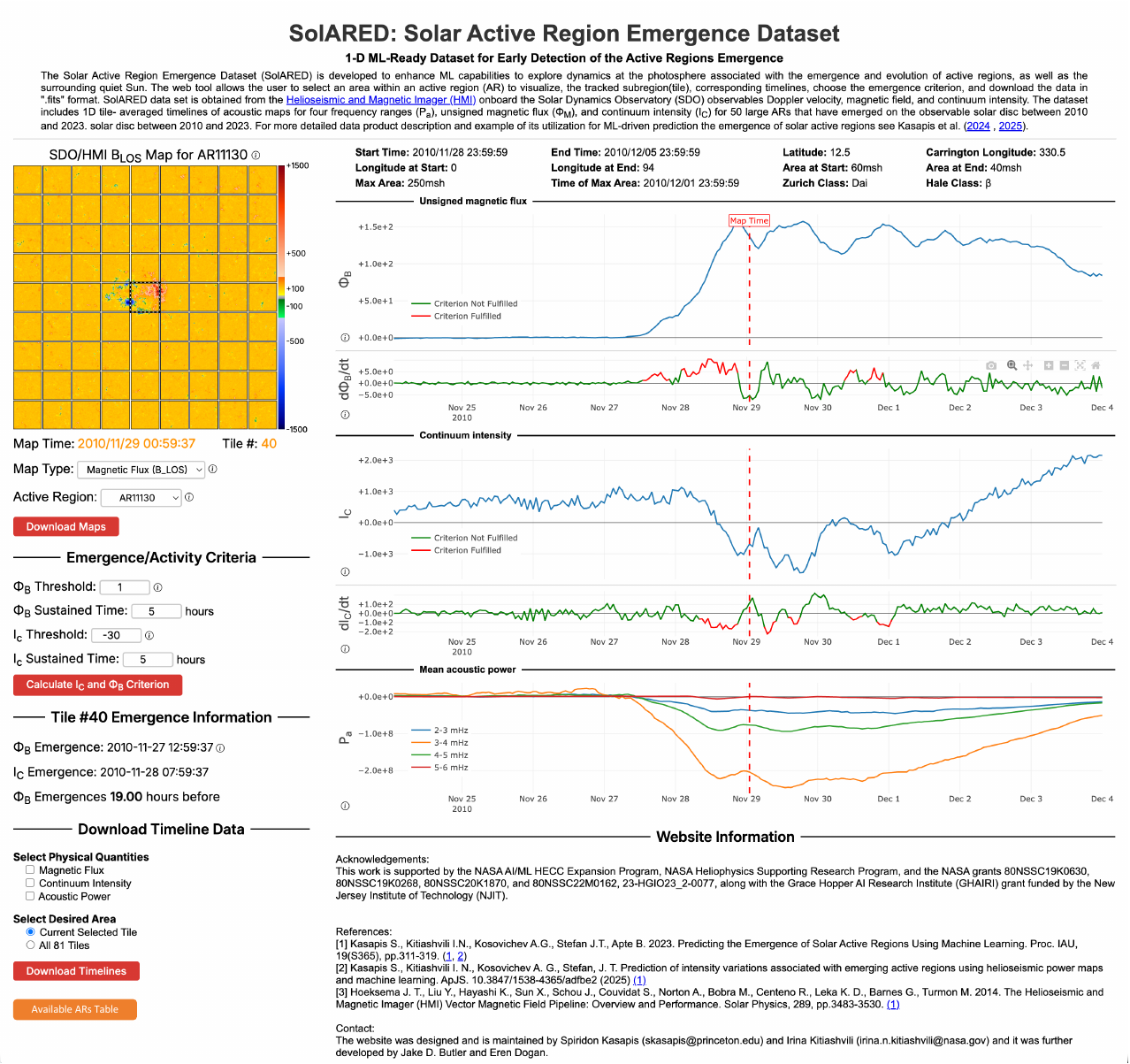} 
    \caption{The interface of the SolARED portal. This web tool has two main components: the data selection panel (left) and the 1D data visualization panel (right).} 
    \label{fig:VizPortal} 
\end{figure}

The front end of the SolARED webpage is split into two sections (Figure \ref{fig:VizPortal}): the data selection panel (left part), which is the main user interface where selections of which AR and tile can be visualized and downloaded, and the data visualization panel (right part), which presents the ML-ready time series data that were produced using the processing steps discussed in this paper.
At the top left of the data selection panel, the tracked $30.66^{\circ} \times 30.66^{\circ}$ area can be viewed, split into a 9-by-9 grid (Figure~\ref{fig:Video}). Using the web interface (Figure~\ref{fig:VizPortal}), the user can select: 
\begin{itemize}
    \item The map of magnetic flux and/or continuum intensity to visualize by clicking on the \textit{Select Map Type} drop-down menu and adjusting the applied color table to highlight the structuring of a region-of-interest in the selected visualization map by dragging the \textit{Color Shift} slider.
    \item The active region from a list of the 50 active regions \citep[Table 4 in][]{kasapis2025prediction} to visualize by clicking on the \textit{Select Active Region} drop-down menu.
    \item By clicking on the map images, the user can select which tile of the 9-by-9 grid should be visualized with the 1D data series on the data visualization panel on the right.
    \item By clicking on any of the three plots of the 1D data visualization panel and moving the red dotted line marked as \textit{Map Time}, the user can change the preselected moment of time for the 2D map plotted on top of the data selection panel.
\end{itemize}

The rest of the data selection panel is split into three parts: \texttt{Emergence Criteria}, \texttt{Tile \# Emergence Information}, and \texttt{Download Data}. Through the \texttt{Emergence Criteria} menu, the user can select the threshold values that define emergence in their plotting and under the \texttt{Tile \# Emergence Information} section. The criteria of an AR emergence start time are defined as the time when a decrease of the continuum intensity time derivative was sustained for more than 3 hours, with a decrease rate over 0.01 in relative units \citep{kasapis2025prediction}. To change properties of the criterion, it is important to keep in mind that the ML-ready data products are in non-dimensional units due to the applied normalization (Section~\ref{sec:flattening}). 

The \textit{Emergence Criteria} section allows a user to customize the emergence criterion based on the time-derivative threshold and the sustained time for the normalized magnetic flux ($d\Phi_B/dt$) and the continuum intensity ($dI_c/dt$). This option will update the color coding to indicate moments of the emergence ($d\Phi_B/dt$ and $dI_c/dt$ in Figure~\ref{fig:VizPortal}). The default definition of emergence criteria is $d\Phi_B/dt < 1$ and $dI_c/dt < -30$, both for longer than 3 hours.

The \texttt{Download Data} section allows the user to specify the data to be downloaded. The user can choose to download the time-series data (labeled `1D') for unsigned magnetic flux, continuum intensity, acoustic power, or all of these physical quantities. The download gets initiated by clicking on the red \textit{Download Timelines} button. The downloaded data files are in the FITS format \citep{wells1981FITS}, which includes a header with parameters describing the observation, such as the start and end times of the tracking interval (T\_START, T\_STOP), the central latitude and longitude of the region (LATHG, LONHG), its spatial extent (WIDTH, HEIGHT), and others. The $\Phi_B$, $I_c$, and $P_A$ time series plots in the 1D data visualization panel can be downloaded as \textit{.png} files by clicking on the camera icon that becomes visible upon hovering over the top right part of a plot (Figure~\ref{fig:VizPortal}). The website also allows the user to download the 2D magnetic flux and continuum intensity maps that are currently being visualized by clicking the \textit{Download Maps} button under the AR selection drop-down menu.

The data visualization panel on the right-hand side of the SolARED webpage includes dynamic plots of the unsigned magnetic flux ($\Phi_B$), continuum intensity ($I_c$), acoustic power ($P_A$), and the time derivatives of the unsigned magnetic flux ($d\Phi_B/dt$) and the continuum intensity ($dI_c/dt$). By dragging across a selected time range in any of the five plots, users can zoom in on the desired interval and adjust the $x$-axis scale accordingly. All five plots are synchronized, so any zoom or range adjustment is reflected across all of them. Hovering over any point on the plots displays the corresponding time and value.
Above the plots, the SolARED interface provides basic information about the selected active region (AR), including its start time (based on the NOAA definition rather than on the emergence criterion, the end time of tracking (when the AR is beyond 75 degrees west from the central merian), the Stonyhurst longitudes and latitude, the Carrington longitude, and the AR area at the NOAA start time, end time, and maximum observed area in the millonths of the Sun's hemisphere. It also lists the time of maximum area, as well as the AR’s Zurich and Hale classifications \citep{chapman1939magnetic, jaeggli2016magnetic}. The list of available active regions and their properties can be found in \cite{kasapis2025prediction} and viewed in a browser by clicking the \textit{Available ARs Table} button at the bottom left of the website. Additional information on how to use the web tool is available by hovering over the ``i" icons located next to various interface elements. 

\section{Case study: AR11158} \label{sec:statistics}



\begin{figure}[h] 
    \centering 
    \includegraphics[width=0.625\textwidth]{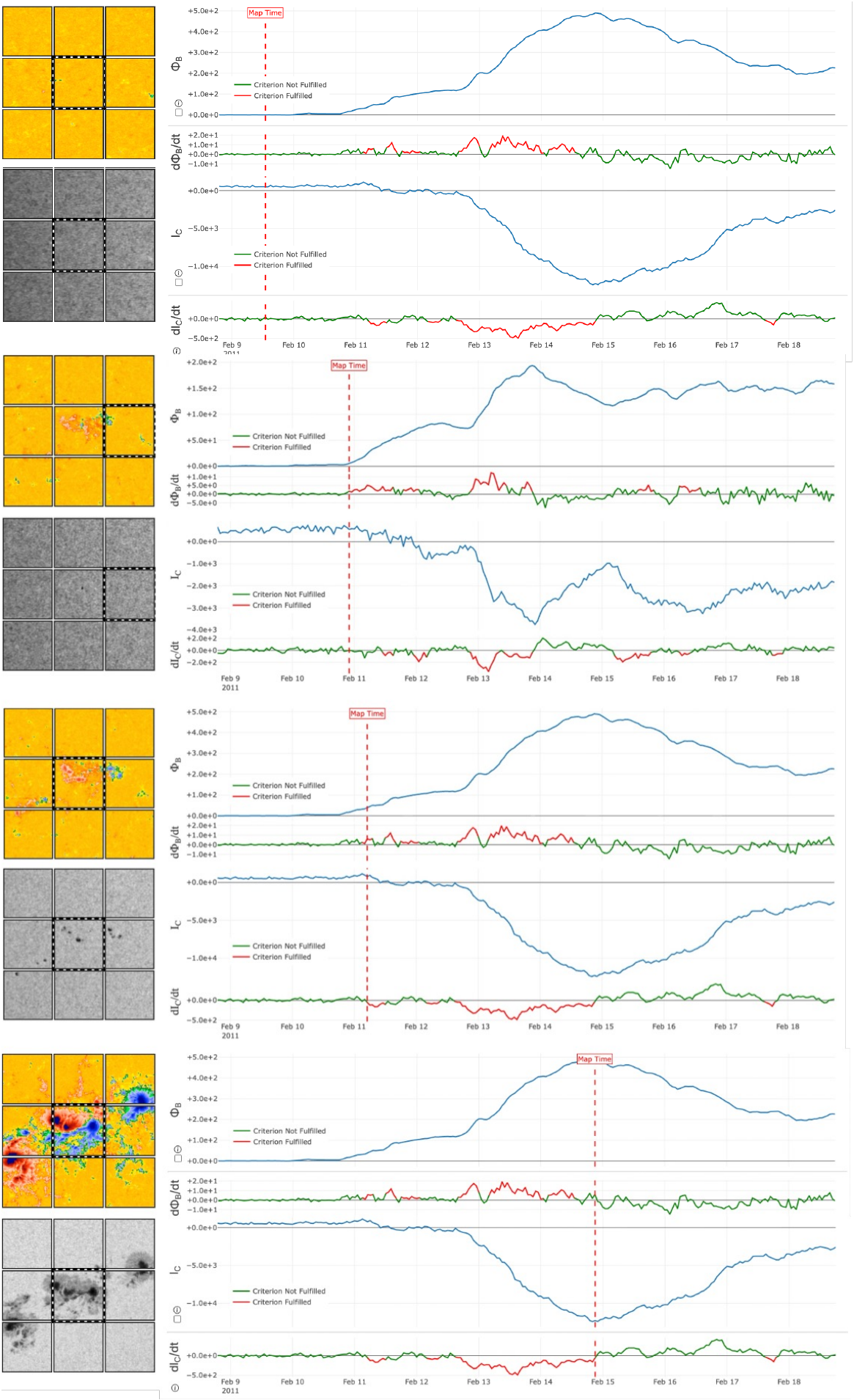} 
    \caption{Visualization of the emergence and evolution of AR 11158 using the SolARED website. The left column shows maps of the 9 centermost magnetic flux (upper color maps) and continuum intensity (grey color-scale maps) tiles of the 9-by-9 grid used in the SolARED dataset. The plots on the right show the corresponding timelines for the selected tiles (dotted outlines on the maps).} 
    \label{fig:evolution} 
\end{figure}

To illustrate the SolARED capabilities, we consider the emergence and evolution of AR 11158. The active region emerged on November 27th, 2010, and persisted on the visible solar disc for seven days until it disappeared behind the West limb on December 4th, 2010. Figure \ref{fig:evolution} (left column) shows maps of the nine centermost tiles in the 9-by-9 grid used in the SolARED analysis, for the magnetic flux (color panels) and the corresponding continuum intensity (grey color-scale panels). Each map shows the AR11158 at different stages of its evolution (from top to bottom): 
\begin{enumerate}
    \item Pre-emergence (quiet sun) phase on 09/02/2011 13:00:00~UT
    \item The first increase in magnetic flux on 10/02/2011 22:00:00~UT
    \item The first decrease in continuum intensity at 11/02/2011 05:00:00~UT
    \item The time when the AR reaches its maximum magnetic flux at 14/02/2011 21:00:00~UT
\end{enumerate}
The times of each phase are denoted using horizontal red dotted lines on the time series plots. Therefore, for AR 11158, based on the default SolARED emergence criteria, the first magnetic flux increase precedes the associated first decrease in continuum intensity by 7 hours, and these changes are observed in the neighboring tiles 40 and 41. At the time the first continuum intensity depletion was observed, the magnetic flux had increased to 10\% of the maximum flux observed throughout the life of the AR, which occurred on 14/02/2011 21:00:00~UT, in tile 40.
The plots on the right show the evolution of magnetic flux, continuum intensity, and their time derivatives for a selected tile, indicated by the black dotted outlines in the left panels.

Using the same analysis for all ARs on the SolARED website, the lead time between the decrease in magnetic flux and the associated decrease in continuum intensity can be quantified. Given the default SolARED website criteria, the first magnetic flux increase precedes the first decrease in continuum intensity in 42 (84\%) out of the 50 ARs. The remaining 15\% of ARs are special cases in which the magnetic flux increases gradually, thereby failing to meet the emergence criterion we have set. On average, the magnetic flux can be observed 16.25 hours ahead of the continuum intensity, with a standard deviation of 18.53 hours, indicating that a diverse population of AR emergence behaviors can be studied using the SolARED dataset.  

%
%

\section{Conclusions} \label{sec:Conclusions}

The SolARED dataset and web portal provide public access to ML-ready data for developing ML models to predict the emergence of ARs \citep{kasapis2023predicting,kasapis2025prediction}.
Using the SolARED website, the user can define their own emergence criteria based on the time derivative of either an SDO/HMI physical quantity. By doing so, the website illustrates the onset of activity within each tile (the green part of the derivative timelines in Figures~\ref{fig:VizPortal} and \ref{fig:evolution} of Sections~\ref{sec:DataPortal} and \ref{sec:statistics}) and therefore the emergence of the entire AR when all tiles are considered. Due to the correction for the solar geometric effect applied, the increase in magnetic flux and decrease in continuum intensity associated with the emergence of each AR are much clearer as observed in Figure~\ref{fig:Correction} of Section~\ref{sec:flattening}.

SolARED delivers to the community machine-learning–ready 1D time series for four observables derived from the SDO/HMI data: the Doppler velocity, continuum intensity, unsigned line-of-sight magnetic flux, and the acoustic power in four frequency ranges. These time series were produced through a standardized four-step processing pipeline: (1) local tracking and remapping of $30.66\times 30.66$ degrees areas in the vicinity of emerging active region, (2) computation of acoustic power maps from Doppler velocity data, (3) spatial downsampling of the resulting 2D time series into a $9\times9$ tile grid, and (4) the data normalization to the standard range of values, $[0,1]$. Thus, SolARED provides uniform, high-quality time series suitable for a wide range of data analysis techniques and ML applications, while preserving the physical signals that enable the detection of changes in the photospheric properties and capture AR emergence before its formation on the surface, as well as the subsequent evolution.

The SolARED website application directly addresses the longstanding gap in publicly available datasets that specifically target the pre-emergence, emergence, and development phases of solar active regions. While previous work has explored various precursors to AR emergence, most existing datasets have not been designed to systematically capture these signals in a machine learning–ready format. By providing uniformly processed time series of continuum intensity, unsigned magnetic flux, and acoustic power for 50 emerging ARs, SolARED enables both physics-based investigations of subsurface-to-surface coupling processes and the development of machine learning approaches for AR emergence prediction \citep[e.g.,][]{kasapis2023predicting,kasapis2025prediction}. In this way, SolARED offers a valuable tool for both observational solar physics and operational space weather forecasting.

In light of the criteria for AI/ML-ready datasets discussed by \cite{2022arXiv220309544N} and \cite{masson2024heliophysics}, SolARED satisfies the essential requirements that distinguish such resources. The dataset is publicly accessible through the SolARED web portal, offers complete, uniformly processed time series of the emerging AR observations, and eliminates the need for additional preprocessing by providing normalized, gap-free data. The standardized pipeline used to create SolARED ensures consistency across all samples, while the documentation and visualization tools enable users to understand and effectively utilize the data. SolARED is a robust starting point for data-driven modeling of emerging ARs, reducing the time-intensive data preparation phase and accelerating the development of machine learning approaches for predicting AR emergence.

\section*{Open Research Section}

The SolARED data portal utilizes Python (Flask) as a backend and plain HTML and javascript as a front-end. Plots and maps are dynamically displayed as JavaScript requests data from Flask and injects the data into the HTML document. The SolARED capabilities and ML-ready data are available for download from \url{https://sun.njit.edu/sarportal/}. The code used to generate and process the dataset, following the methodology described in the paper, is available at \url{https://github.com/skasapis}. 

\acknowledgments

We want to thank the NAS Visualization Team (Nina McCurdy, Timothy Sandstrom, and Christopher Henze) for their help with this project’s visualizations. The computational resources are provided by the NASA Ames Research Center Supercomputing Facility. This work is supported by the NASA AI/ML HECC Expansion Program, and the NASA grants 23-HGIO23\textunderscore2-0077, 20-HSR20\textunderscore2-0037, 80NSSC19K0630, 80NSSC19K0268, 80NSSC20K1870, and 80NSSC22M0162. 

%
%

\bibliography{biblio}

\end{document}